\documentclass[%
 reprint,
superscriptaddress,
 amsmath,amssymb,
 aps,
prb,
twocolumn
]{revtex4-2}

\usepackage{graphicx,bm,braket,color,hyperref,comment}
\usepackage{fancybox}
\usepackage{ulem}
\hypersetup{colorlinks=true,citecolor=blue,linkcolor=blue,urlcolor=blue}
\newcommand{\bq}{ \bm{q} }
\newcommand{\bk}{ \bm{k} }

\begin{document}


\title{Order parameter fluctuation effects on current-induced magnetization
}

\author{Genta Furuya}

\author{Kazumasa Hattori}%
\affiliation{Department of Physics, Tokyo Metropolitan University, 1-1, Minami-osawa, Hachioji, Tokyo 192-0397, Japan}

\date{\today}

\begin{abstract}
We investigate the impact of order parameter fluctuations on magnetoelectric effects in metallic systems using classical Monte Carlo simulations. We focus on a a chiral quadrupole order in a distorted kagome lattice in a model incorporating conduction electrons and classical orbital moments.  The ordered orbital moments  break mirror symmetry and couple with the conduction electrons, leading to a current-induced magnetization driven by the quadrupole order. Our findings reveal that order parameter fluctuations significantly affect the current-induced magnetization, strongly suppressing the response over a broad temperature range below the transition temperature. Additionally, we analyze the influence of Fermi surface properties and associated matrix elements on this phenomenon. Our results highlight the crucial role of fluctuation effects, demonstrating a qualitatively distinct temperature dependence of the current-induced magnetization compared to that of the order parameter itself.
\end{abstract}

\pacs{Valid PACS appear here}
\maketitle



\section{Introduction} \label{sec:Intro}
Cross-correlations, such as magnetoelectric effects \cite{Fiebig2005-kg}, play a crucial role in next-generation technologies and have been actively explored in recent years. Off-diagonal response functions provide opportunities to control functional degrees of freedom using non-conjugate fields. For instance, noncentrosymmetric materials offer potential applications in the electric control of magnetic moments and its inverse, magnetic control of polarizations \cite{Khomskii2006}. These phenomena have been intensely studied, particularly in the field of spintronics \cite{Hoffmann2015-tb} and various multiferroic materials \cite{Tokura2014,Fiebig2016-or}. Such cross-correlations between different degrees of freedom have been systematically analyzed using group-theoretical classifications, including multipole theory \cite{Hayami2018,Kusunose2020}, magnetic point groups \cite{Yatsushiro2021_classification}, and spin space groups \cite{Smejkal2022-fi,Liu2022-kg,Watanabe2024-zr}.

Once the symmetry of a material or phase is identified, the emergence of each cross-response is determined by its symmetry. This is a key advantage of symmetry classification, providing a guideline for designing materials with functional properties related to cross-correlated phenomena. Recently, magnetoelectric effects in metals have attracted considerable attention, particularly following the discovery of the Edelstein effect \cite{Edelstein1990} in elemental tellurium (Te) \cite{Furukawa2017}. The crystal structure of Te lacks both inversion and mirror symmetries, leading to a finite current-induced magnetization (CIM). This effect has been successfully observed through sophisticated nuclear magnetic resonance experiments.

More recently, UNi$_4$B has been found to exhibit the Edelstein effect in its symmetry-broken phase below the magnetic ordering temperature $T_c$ \cite{Saito2018}. However, the anisotropy of the observed current-induced magnetization (CIM) contradicts the expected behavior for the known magnetic toroidal order in this triangular lattice compound \cite{Mentink1995-st, Hayami2014}. The anisotropy in the magneto-current tensor can serve as a tool for identifying the order parameter, and an alternative possible order parameter for UNi$_4$B has been theoretically proposed \cite{Ishitobi2023-ep}. A key aspect of this system is the presence of two distinct contributions to the induced magnetic moments: one driven by the current and the other by an applied electric field. While the proposed order parameter can explain the anisotropy of the CIM, the temperature ($T$) dependence of the CIM below $T \leq T_c$ remains poorly understood. In UNi$_4$B, the absolute value of the CIM increases rapidly with a concave-upward $|T_c - T|$ dependence \cite{Saito2018}. Even near $T_c$, the temperature range in which the CIM follows the order parameter—typically behaving as $\sim \sqrt{T_c - T}$—is quite narrow. A similar behavior is observed in Ce$_3$TiBi$_5$, where the $T$-dependence of the CIM in its magnetically ordered state is almost linear in $T_c - T$ near $T_c$ \cite{Shinozaki2020}. This temperature dependence differs significantly from that of magnetoelectric susceptibility observed in well-known systems such as Cr$_2$O$_3$ \cite{Folen1961-cb}. These observations naturally lead to considerations regarding the role of the Fermi surface and fluctuation effects in the order parameters, which we analyze in detail in this study.

\begin{figure}[t]
\centering
\includegraphics[width=0.48\textwidth]{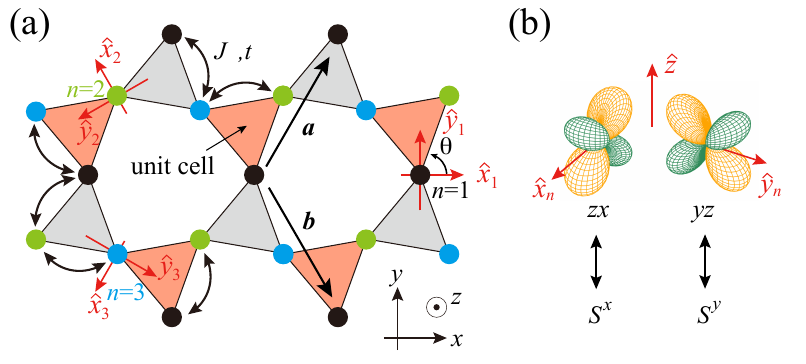}
\caption{(a) Distorted kagom\'e structure with $\theta\ne \pi/3$. Sublattice sites ($n=1,2,3$) are shown by different colors, which form a unit cell of orange triangle. Red arrows are the local coordinates $\hat{x}_n$ and $\hat{y}_n$. Primitive translational vectors $\bm{a}$ and $\bm{b} $ are also indicated by black arrows. Double-headed arrows represents the exchange interaction path and hopping elements. (b) Schematic picture of orbital degrees of freedom $\{yz,zx\}$ and its correspondence to $\bm{S}$.}
\label{fig:struc}
\end{figure}

In this paper, we focus on a chiral quadrupole order of $\{yz,zx\}$ type in the distorted kagom\'e structure as shown in Fig.~\ref{fig:struc}. This order breaks the all mirror symmetries, leading to a longitudinal CIM, i.e., magnetization parallel to the current \cite{Ishitobi2025-uq}. Recently, such nonmagnetic orders have been identified in URhSn \cite{Chevalier1995-mp,Shimizu2020,Maurya2021} as evidenced by the resonant x-ray \cite{Tabata_JPS} and nuclear magnetic resonance experiments \cite{Tokunaga_JPS,Kusunose2023-de}. In contrast to the magnetoelectric effects due to magnetic orders, the effects caused by electric quadrupole orders consist solely of a current-induced contribution. This simplifies the theoretical analysis, as it does not involve additional electric-field-induced terms.

This paper is organized as follows. In Sec.~\ref{sec:orb}, we show the classical orbital model consisting of the $\{yz,zx\}$ type quadrupole moments and briefly explain the classical Monte Carlo results. In Sec.~\ref{sec:CIM}, we analyze CIM for a classical Kondo lattice model by taking into account itinerant electrons in addition to the classical orbital model used in Sec.~\ref{sec:orb}. The analysis is hybrid one,  
where we carry out the exact diagonalization of the itinerant model under a large number of snapshot configurations of the classical orbital moments. 
Section~\ref{sec:dis} is dedicated to discussions on the microscopic mechanism of CIM in the presence of fluctuating order parameters. Finally, Sec.~\ref{sec:sum} provides a summary of our study.

\section{Classical orbital model}\label{sec:orb}
In this section, we introduce a minimal orbital model which can describe the chiral quadrupole order in the distorted kagom\'e structure \cite{Ishitobi2025-uq}. The model is analyzed by the classical Monte Carlo simulation. These become the basis for the analysis for the CIM in Sec.~\ref{sec:CIM}.  

\subsection{Model}
Let $\bm{S}_i=(S_i^x,S_i^y)$ be the orbital degrees of freedom at the site $i$ with $\sim \{zx,yz\}$ type in the local coordinates, $\hat{x}_n$ and $\hat{y}_n$, as shown in Fig.~\ref{fig:struc}. We analyze a simple orbital model which can capture the essential nature of chiral quadrupole order. We use the following interacting orbital Hamiltonian 
\begin{align}
	H_{\rm cl}&=J\sum_{\langle i,j\rangle} \bm{S}_i\cdot \bm{S}_j
	+\delta\sum_i \Big[(S_i^x)^2-(S_i^y)^2\Big]
,\label{eq:Hclass}
\end{align}
where $J$ is the coupling constant on the nearest-neighbor $ij$ bonds. The second term proportional to $\delta$ is the local anisotropy potential which reflects the inequivalence between the local $\hat{x}_n$ and $\hat{y}_n$ directions. 

We also make another approximation that the vector $\bm{S}_i$ takes four discrete values $(\pm 1,0)$ and $(0,\pm 1)$. We have carried out classical Monte Carlo (MC) simulations of this simplified model in the distorted kagom\'e structure. We use Metropolis algorithm combined  with Wolff's cluster updates \cite{Wolff}. The thermal average of observables $A$, $\langle A \rangle$, is calculated by averaging typically over $5\times 10^6$ MC steps.  Since our main purpose is to analyze current-induced magnetization under the chiral orders in Sec.~\ref{sec:CIM}, we will not analyze its critical properties and details of the model for large system sizes. The universality class is the two-dimensional Ising class as is evident from the anisotropy. The purpose of this section is to obtain the qualitative view of the phase diagram of classical fields used in the Kondo lattice model in Sec.~\ref{sec:CIM}. 

\begin{figure}[t]
\centering
\includegraphics[width=0.48\textwidth]{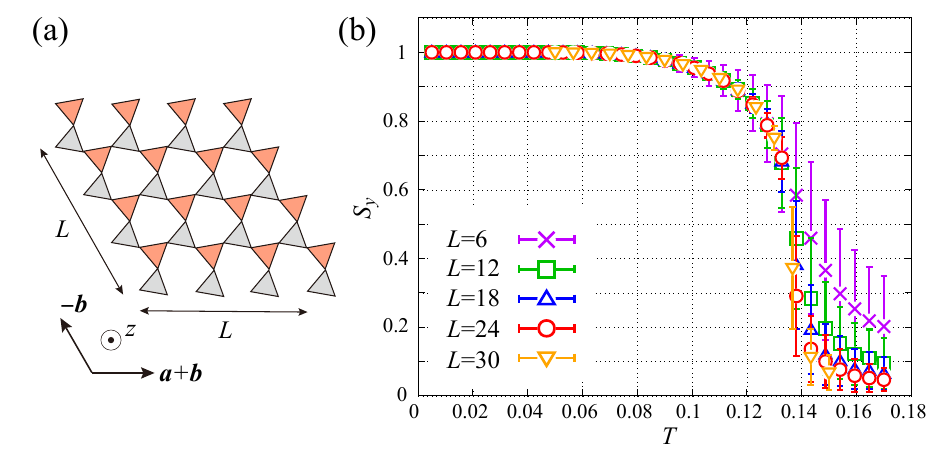}
\caption{(a) Cluster shape of distorted kagom\'e structure used in the simulations ($L=4$). The system contains $3L^2$ sites with the periodic boundary conditions along the ${\bm{a}+\bm{b}}$ and $\bm{b}$ directions. (b) Temperature ($T$) dependence of the order parameter $S_y$ for $J=-0.1$, $\delta=-0.02$, $L=6,12,18,24$, and $30$, where the transition temperature is $T_c\simeq 0.135$.}
\label{fig:Oyz}
\end{figure}

\begin{figure}[t]
\centering
\includegraphics[width=0.48\textwidth]{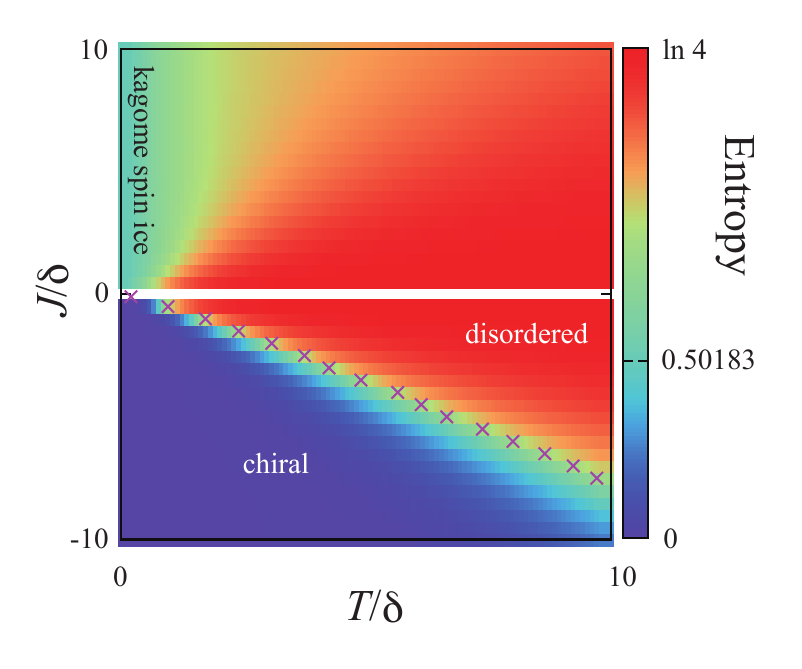}
\caption{$T$--$J$ phase diagram for $\delta>0$ and $J^z=0$. Color represents the entropy per site calculated by the $T$-integration of the specific heat divided by $T$. $\times$ is the transition temperature of the chiral phase estimated by crossing points of Binder ratio. For $J>0$, the kagom\'e spin ice configurations are realized and the spin-ice entropy 0.50183 \cite{Kano1953-dn} is reproduced.}
\label{fig:phase}
\end{figure}

\subsection{Numerical results}
Throughout this paper, the numerical simulations have been done in a parallelogram cluster whose edge length is $L$ and in the periodic boundary conditions along the direction of the primitive translational vectors $\bm{a}+\bm{b}$ and $\bm{b}$ shown in Fig.~\ref{fig:Oyz}(a). For $J<0$ and $\delta>0$, the system undergoes a phase transition as lowering temperature ($T$). The order parameter $S_y$ is calculated by $S_y\equiv N^{-1}\sum_i\langle |S_i^y|\rangle$, where $N=3L^2$. Remember that we use the local axes as shown in Fig.~\ref{fig:struc}. A typical $T$ dependence of the order parameter is shown in Fig.~\ref{fig:Oyz}(b) for $L=6,12,18,24$, and 30. The transition temperature $T_c$ is determined by the scale invariant point of the Binder ratio and the result is $T_c\simeq 0.135$ for $J=-0.1$. Here, although we can carry out classical model simulations for larger system sizes, we show the data only for $L\le 30$. This is because we will use them for the analysis of the Kondo lattice model in Sec.~\ref{sec:CIM} with much heavier computational cost. For $J<0$ and $\delta<0$, the order parameter is the $x$ component $S_x$ and the results are similar to the cases for $\delta>0$ and will not be discussed hereafter. 

Figure \ref{fig:phase} shows $T$--$J$ phase diagram for $\delta>0$. The color represents the entropy calculated by integrating the specific heat divided by $T$ from the lowest $T$ to sufficiently high $T$ under the assumption that the entropy at high $T$ saturates $\ln 4$. The anisotropy term in Eq.~(\ref{eq:Hclass}) favors the  $\bq=\bf{0}$ uniform order with $S_y\ne 0$ for $J<0$. For antiferroic interaction $J>0$, there remains finite residual entropy of the kagom\'e spin ice $\sim 0.50183$ per site \cite{Kano1953-dn}. Note that owing to $\delta>0$, $\bm{S}_i=(\pm 1,0)$ components are frozen out at around $T\sim \delta$ and the low temperature physics is Ising-like.  

\section{Current induced magnetization}\label{sec:CIM}
In this section, we analyze the CIM in the chiral quadrupole ordered phase discussed in Sec.~\ref{sec:orb}. The CIM was discussed many years ago in  a noninteracting metal with nonmagnetic impurities \cite{Levitov1985,Inoue2003-vh} and Fermi liquid theory \cite{Fujimoto2007-lb}, and dynamical mean-field theory \cite{Peters2018-ad}. We here analyze strong order parameter fluctuations in a noninteracting metal. We will introduce a classical Kondo lattice model where the classical localized moments are $\{zx,yz\}$ type orbital  moments. The CIM is analyzed in the framework of the Kubo theory for response functions \cite{Kubo_1957}. The temperature dependence of the CIM is calculated by averaging the MC snapshot  configurations of the orbital moments for each temperature.

\subsection{Classical Kondo lattice model}
We now introduce a toy two-orbital nearest-neighbor tight-binding model on the distorted kagom\'e structure with the classical localized moment ${\bm S}_i=(S_i^x,S_i^y)$: 
\begin{align}
	H &= \sum_{i,j}  \psi^\dagger_i \hat{t}_{ij}\psi_j 
	+ K\sum_{i} \sum_{\nu=x,y}\psi^\dagger_i \hat{\tau}_y \hat{\sigma}_\nu \psi_i  S^\nu_i
	\label{eq:conduction}
,\end{align}
where $\psi_i^\dag$ $=$ $(c^\dag_{i,+\uparrow}, c^\dag_{i,+\downarrow}, c^\dag_{i,-\uparrow}, c^\dag_{i,-\downarrow})$ is the creation operator of the electron with the orbital (spin) $\lambda$$=$$\pm$ $ (\sigma=\uparrow,\downarrow)$. $\tau(\sigma)$'s are the Pauli matrices for the orbital (spin) sector in each local coordinate $\{\hat{x}_n,\hat{y}_n\}$ shown in Fig.~\ref{fig:struc}(a). 
We assume the electrons are  
$p$-electron-like with $\{ c_{i,+\sigma}, c_{i,-\sigma} \}$=$\{ c_{i,x\sigma}, c_{i,y\sigma} \}$ as a representative example. As a minimal model which possesses distorted kagom\'e structure, we take into account hopping matrix elements between the nearest-neighbor (NN) sites via the $(pp\sigma)$-type orbital-dependent hopping $t_{pp\sigma}=1$. See, Fig.~\ref{fig:struc} and their expressions in Appendix \ref{app:hop}. 
The specific orbital character is not important as long as they can couple with the chirality order. For example, they can be $d$-orbital $\{ d_{i,x^2-y^2\sigma}, d_{i,xy\sigma} \}$ or  $\{ d_{i,zx\sigma}, d_{i,yz\sigma} \}$. The second term represents the orbital Kondo coupling $K$. Note that the $\hat{\tau}^y$ transforms as the $z$ component of magnetic dipole moment. In actual calculations, we add the chemical potential term $-\mu\sum_i\psi^\dagger_i\psi_i$ to Eq.~(\ref{eq:conduction}) in the following in order to tune the electron filling $n_{\rm el}$.

\begin{figure}[t!]
\centering
\includegraphics[width=0.5\textwidth]{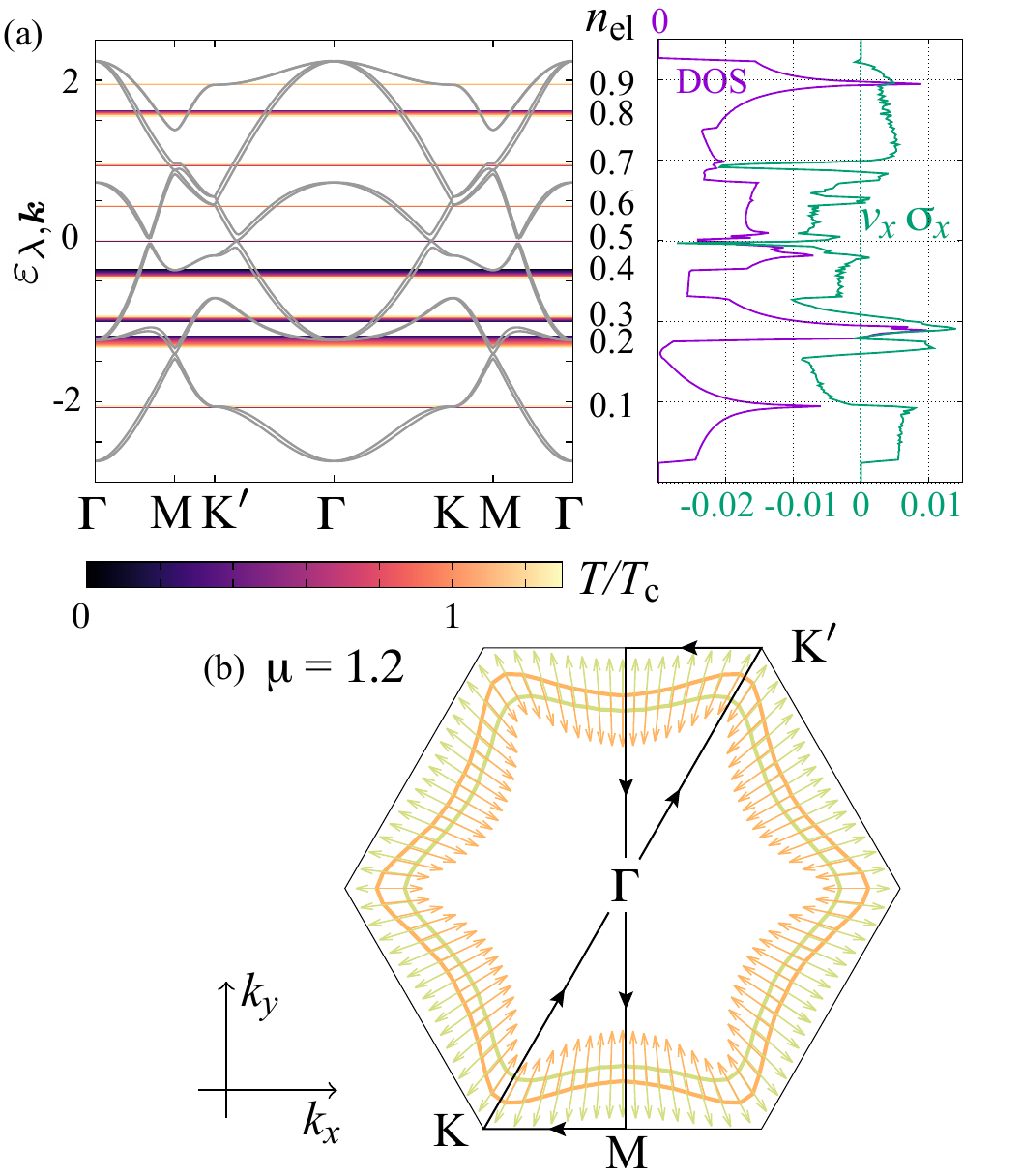}
\caption{(a) $\varepsilon_{\bk,n}$ along the path $\Gamma$-M-K-$\Gamma$-K$^\prime$-M-$\Gamma$ (left) and the DOS (purple) and the expectation value of $\hat{v}_x\hat{\sigma}_x$ denoted as $v_x\sigma_x$ (green) for the Bloch state (right) for $t_{pp\sigma}=1$, $\mathcal K=-0.1$, and $\theta=75^\circ$. $T$ dependence of $\mu$ for several $n_{\rm el}$'s are shown by horizontal lines with the colors representing $T/T_c$, where $T_c\simeq 0.135$ for $J=-0.1$ and $\delta=0.02$. (b) Fermi surfaces for $\mathcal K=-0.1$ and $\mu=1.2$. The Bloch band spin moments are indicated by arrows at each FS with the same color (orange and lime). The path corresponding to the horizontal axis in (a) is also shown.}

\label{fig:FermiSurface}
\end{figure}

\subsection{Uniform orders}
It is instructive to see the band structure for the chiral quadrupole order $S_i^y={\mathcal C}=$ const., where the system has the translational symmetry and the wave number ${\bm k}$ is a good quantum number. The one-particle energy $\varepsilon_{\lambda,{\bm k}}$ of the model (\ref{eq:conduction}) exhibits spin-split Fermi surfaces (FS's) owing to the lack of mirror symmetries and the effective spin-orbit coupling proportional to ${\mathcal C}$, where $\lambda$ is the band index. The left panel of Fig.~\ref{fig:FermiSurface}(a) shows the energy dispersion $\epsilon_{\lambda,\bm{k}}$ for ${\mathcal K}\equiv K{\mathcal C}=-0.1$. The density of states (DOS) and the expectation value of $\hat{v}_x\hat{\sigma}_x$ for the Bloch state are also shown as a function of the energy. See Eq.~(\ref{eq:def_C}) for more general definition. Here, $\hat{v}_x$ is the $x$ component of the velocity (current) operator. See the definition in Appendix~\ref{app:current}. There are several sharp peaks and kinks in both quantities, which reflect the band bottom/top and the van Hove singularities. We will come back to these aspects when discussing the $T$ dependence of the CIM in Sec.~\ref{sec:CIM_result}.   The horizontal lines represent $T$ dependence of the chemical potential $\mu$ for each electron number $n_{\rm el}$ indicated. Here, $T$ dependence of the order parameter $C=S_y$ for $L=24$ in Fig.~\ref{fig:Oyz}(b) is assumed. In Fig.~\ref{fig:FermiSurface}(b), the FSs for $\mathcal{K}=-0.1$ are shown.
The arrows indicate the spin moments of the Bloch state at ${\bm k}$. The radial spin textures are hedgehog (lime) and antihedgehog (orange) types for $C\ne 0$. The spin-split band in their dispersion clearly shows that the chiral order parameter $C$ induces the chirality of the itinerant electrons. 

\subsection{Kubo formula }

One of significant properties arising in chiral materials is the Edelstein effect \cite{Edelstein1990}: current-induced magnetization due to the spin polarization on the FS's. The magnetoelectric coefficient tensor $\alpha_{ij}$ is  defined by $m_i = \sum_{j}\alpha_{ij}E_j$ with the magnetization $m_i$ and electric field $E_j$ ($i,j=x,y,z$). In the present system, although the inversion symmetry is already broken without the chiral order ${\mathcal C}$$=$$0$, there is no Edelstein effect in the distorted kagom\'e system with the D$_{3h}$ point group.

In general, $\alpha_{ij}$ can be obtained from the current-magnetization correlation function $\chi_{ij}$ as \cite{Kubo_1957}
\begin{align}
&\alpha_{ij}=\lim_{\omega\to 0}\frac{\chi_{ij}(\omega)-\chi_{ij}(\omega=0)}{i\omega},\label{eq:alpha}\\
	&\chi_{ij}(\omega)=\frac{1}{N^2}\sum_{{\bm{r},\bm{r}',n,n'}}\int_{-\infty}^\infty dt \chi_{ij}(t,\bm{r},n,\bm{r}',n') e^{i\omega t},\label{eq:chi_w}\\
	&\chi_{ij}(t,\bm{r},n,\bm{r}',n')=i\langle [\hat{\sigma}_i(t,\bm{r},n),\hat{v}_j(0,\bm{r}',n')]\rangle\theta(t).\label{eq:chi_t}
\end{align}
Here, $\sigma_i(t,\bm{r},n)$ is the $i$th component of the spin at the time $t$, at the unit cell position $\bm{r}$ with the sublattice index $n$. For simplicity, we do not discuss the orbital part and concentrate on the spin: $m_i=\sigma_i$.
Similarly, $v_j$ is the $j$th component of the velocity (current). See Appendix \ref{app:current}. $\theta(t)$ is the step function. Note that $i$ and $j$ indicate the direction measured by the global coordinates, i.e., $\hat{x}_1, \hat{y}_1$, and $\hat{z}$.

When the system is translationally invariant, $S_i^\nu=C$, one can use the Bloch basis and 
$\alpha_{ij}$ for nonmagnetic orders is given by \cite{Hayami2018} 
\begin{align}
\alpha_{ij}(T) &= \frac{1}{\eta} \frac{1}{N}\sum_{\lambda,{\boldsymbol k}} \sigma_{i\lambda,{\bk}}v_{j\lambda,{\bm k}} \left(-\frac{\partial f_{\lambda,{\boldsymbol k}} }{\partial \varepsilon_{\lambda,{\bm k}}}\right)
\label{eq:response}
.\end{align}
Here, we have introduced 
a phenomenological damping rate $\eta$. The $i$th component of the velocity and the spin for the band $\lambda$ at $\bk$ are 
$v_{i\lambda,{\bm k}}
$ and $\sigma_{i\lambda,\bk}$, respectively. $f_{\lambda,{\bm k}}$ is the Fermi distribution function $f(\epsilon_{\lambda,\bm{k}})$ with the temperature $T$ and the energy $\varepsilon_{\lambda,{\bm k}}$ for the model (\ref{eq:conduction}). Note that this is so called Fermi surface term and inter band contributions are absent for nonmagnetic orders \cite{Watanabe2018}.

The expression (\ref{eq:response}) is useful when one try to capture the qualitative behavior (e.g., finite or zero) of $\alpha_{ij}$. Our purpose is to analyze the order parameter fluctuation on $\alpha_{ij}(T)$ and Eq.~(\ref{eq:response}) is not applicable. To this end, we calculate $\alpha_{ij}(T;\{\bm{S}(T)\})$ for a given snapshot configuration $\{\bm{S}(T)\}$ at temperature $T$ obtained in the Monte Carlo simulation for the classical orbital model (\ref{eq:Hclass}). We average $N_{\rm MC}=1000$ sets of $\alpha_{ij}(T;\{\bm{S}(T)\})$'s to obtain $\alpha_{ij}(T)$, which can be regarded as an expectation value of $\alpha_{ij}$ which includes the effects of (thermal) order parameter fluctuations. To calculate $\alpha_{ij}(T;\{\bm{S}(T)\})$, we numerically diagonalize the itinerant electron Hamiltonian (\ref{eq:conduction}) with snapshot configuration $\{\bm{S}(T)\}$ and we obtain the single-particle energy $\epsilon_{a}(\{\bm{S}(T)\})$ and the wavefunction $|a;\{\bm{S}(T)\}\rangle$ ($a=1,2,\cdots 12N$). They are sufficient to calculate Eqs.~(\ref{eq:alpha})--(\ref{eq:chi_t}). The final expression is 
\begin{align}
	\alpha_{ij}(T)&=\eta\left\langle
	\sum_{a,a'} \frac{F_{aa'}}{N^2}\!\!\!\sum_{\bm{r}n,\bm{r}'n'}\!\!\!\!
	\frac{\sigma_{aa'}^i(\bm{r},n) v_{a'a}^{j}(\bm{r}'n')}{(\epsilon_{a}-\epsilon_{a'})^2+\eta^2}
	\right\rangle_{\!\!\!\bm{S}(T)}\!\!\!\!, \label{eq:alpha_Av}
\end{align}
where $\left\langle \cdot \right\rangle_{\bm{S}(T)}$ indicates the average over $N_{\rm{MC}}$ configurations randomly chosen in the thermalized snapshot in the MC simulation and 
\begin{align}
	&F_{aa'}=-\frac{f(\epsilon_a)-f(\epsilon_{a'})}{\epsilon_{a}-\epsilon_{a'}}\simeq -\frac{\partial f(\epsilon_a)}{\partial \epsilon_a}\ {\rm for}\ \epsilon_a=\epsilon_{a'},\\
&\sigma_{aa'}^i(\bm{r},n)=	\langle a;\{\bm{S}(T)\}|\hat{\sigma}^i(0,\bm{r},n)|a';\{\bm{S}(T)\}\rangle,\\
&v_{aa'}^i(\bm{r},n)=	\langle a;\{\bm{S}(T)\}|\hat{v}^i(0,\bm{r},n)|a';\{\bm{S}(T)\}\rangle.
\end{align}
In the chirality order phase, only the longitudinal responses $\alpha_{xx}=\alpha_{yy}$ are finite, which is related to the hedgehog spin texture for finite $C$ as shown in Fig.~\ref{fig:FermiSurface}(b).

\subsection{Numerical results}\label{sec:CIM_result}
Now, we discuss the numerical results of the CIM. We fix the parameters $J=-0.1$, $\delta=-0.02$, $\theta=75^\circ$, and $t_{pp\sigma}=1$. This single choice of the parameter set with the transition temperature $T_c\simeq 0.135$ as shown in Fig.~\ref{fig:Oyz}(b) is sufficient for analyzing the fluctuation effects on the CIM by varying the electron filling $n_{\rm el}$. Since transverse responses $\alpha_{ij}(T)$ with $i\ne j$ vanish even for $T<T_c$, we discuss the longitudinal one $\alpha_{xx}(T)=\alpha_{yy}\equiv \alpha(T)$. Before showing the data, we first list the quantities we calculated and their notations: 
\begin{align}
	\alpha(T,L):& \ {\rm{via\ Eq.}}~(\ref{eq:alpha_Av})\ {\rm for\ the\ system\ size}\ L. \nonumber\\
	\alpha_{\rm uni}(T,L):& \ {\rm via\ Eq.}~(\ref{eq:alpha_Av})\ {\rm with\ a\ uniform}\ S_y \ {\rm given \ by \ the} \nonumber\\ 
	& {\rm classical \ model \ (\ref{eq:Hclass}) \ 
	for\ the\ system\ size}\ L.\nonumber\\	
	\alpha_{\rm uni}^{(\beta)}(T,L):& \ {\rm via\ Eq.}~(\ref{eq:alpha_Av})\ {\rm with\ a\ uniform}\ S_y=t^\beta\theta(t), \nonumber\\
	& {\rm where}\ t=(T_c-T)/T_c.\nonumber\\		
	\alpha^{\bm{k}}_{\rm uni}(T,L):& \ {\rm{via\ Eq.}}~(\ref{eq:response})\ {\rm with\ a\ uniform}\ S_y \ {\rm given \ by \ the} \nonumber\\ 
	& {\rm classical \ model \ (\ref{eq:Hclass})}.\nonumber
\end{align}
Note that $\alpha(T,L)$ includes the effects of the order parameter fluctuations, $\bm{S}_i\ne \bm{0}$ in a snapshot $\{\bm{S}(T)\}$, while $\alpha_{\rm uni}(T,L)$, $\alpha_{\rm uni}^\beta(T,L)$, and $\alpha_{\rm uni}^{\bm{k}}(T,L)$ do not. 
For $\alpha_{\rm uni}^{\bm{k}}(T,L)$, we use the data of $S_y$ for finite $L$ and a sufficiently fine $\bm{k}$ mesh for the summation in the first Brillouin zone, while for $\alpha_{\rm uni}$ and $\alpha_{\rm uni}^{(\beta)}$ we carry out the summation in the real space for finite system size $L$. Thus, we can regard $\alpha_{\rm uni}^{\bm{k}}(T,L)\simeq \lim_{L\to \infty}\alpha_{\rm uni}(T,L)$ at low temperature where $S_y\simeq 1$ even for finite $L$. Since the MC snapshots $\{\bm{S}(T)\}$ include two opposite domains with $S_y>0$ and $S_y<0$, we fix them so that $S_y> 0$ by changing the sign of $S^y_i$ for the all $i$. The uniform order parameter for $\alpha_{\rm uni}(T)$ and $\alpha_{\rm uni}^{\bm{k}}(T)$ is given by the data for $L=24$ shown in Fig.~\ref{fig:Oyz}(b) and $S_x=0$. The data will be presented in the following show finite values even for $T>T_c$ but they are due to the finite-size effects and the above domain-fixing procedure. Thus, we concentrate on $T\le T_c$ and will not discuss the disordered phase.  

\begin{figure}[t!]
\centering
\includegraphics[width=0.48\textwidth]{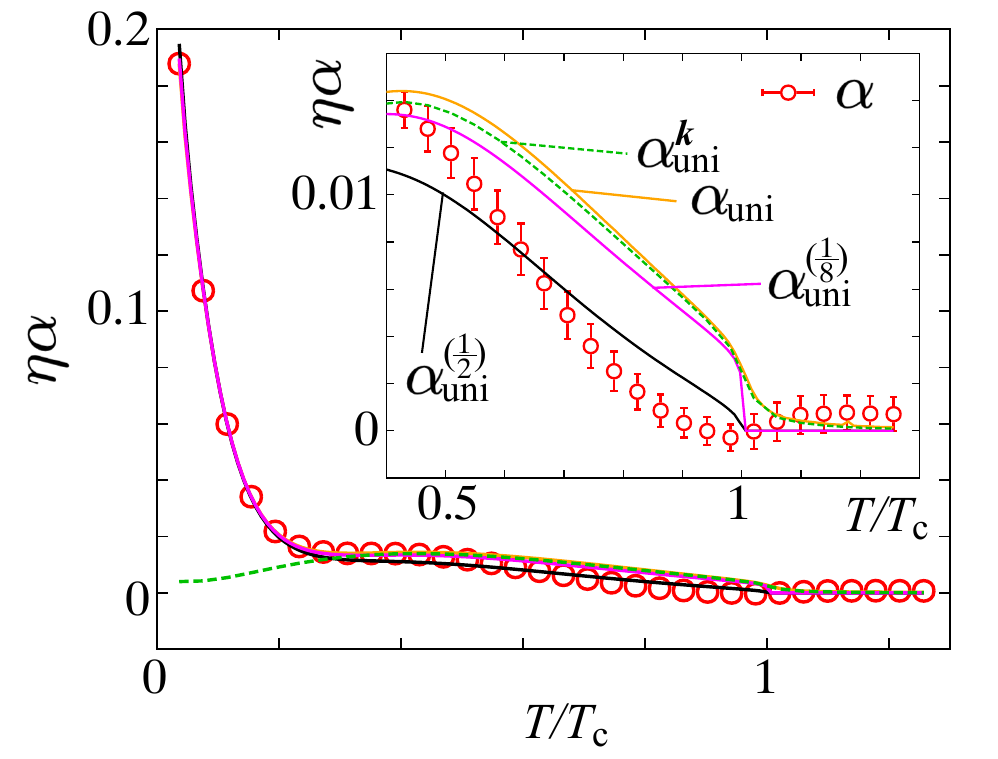}
\caption{$\eta\alpha(T,L=24)$ (open circle) as a function of $T$ with $\eta =0.5/L^2\simeq 0.00086$ and $n_{\rm el}=0.2$. Lines represent various type of $\alpha_{\rm uni}$'s. $\alpha^{\bm k}_{\rm uni}$ is calculated by using $S_y$ for $L=24$. $\alpha_{\rm uni}^{(1/2,1/8)}$ corresponds to the mean-field and two-dimensional Ising type variation of $S_y$, both of which are phenomenological one extrapolated to $T=0$. $\alpha,\alpha_{\rm uni}$, and $\alpha^{\bm{k}}_{\rm uni}$ are finite even for $T>T_c$ owing to the finite-size effect. 
}
\label{fig:alpha_1}
\end{figure}

Figure \ref{fig:alpha_1} shows $T$ dependence of $\eta\alpha(T,L=24)$ (open circle) and various types of $\eta\alpha_{\rm uni}(T,24)$'s (lines) for $n_{\rm el}=0.2$. Here, we show $\eta \alpha$, since $\eta \alpha$ becomes $\eta$ independent when the system has translational invariance as is evident from Eq.~(\ref{eq:alpha}).  The value of $\eta$ is fixed as $\eta L^2=$ const. This parameterization is natural since the finite-size energy gap is roughly proportional to [band width]/[numbers of states] for finite $L$. In Fig.~\ref{fig:alpha_1}, we set to $\eta L^2=0.5$. 
Since there is non-negligible finite-size gap between neighboring energy eigenvalues for $L=24$, the data for low temperatures are not valid as is evident from the fact that $\chi^{\bm k}_{\rm uni}$ deviates from the others. Here, the former should be correct near $T=0$. Thus, we concentrate on the data for $T\gtrsim 0.4T_c$ as shown in the inset of Fig.~\ref{fig:alpha_1}. First, we notice that $\alpha(T,24)$ is strongly suppressed near $T\simeq T_c$, comparing it to $\alpha_{\rm uni}$'s. This suggests that the CIM is suppressed by the order parameter fluctuations near $T\simeq T_c$. Furthermore, the $T_c-T$ dependence is rather concave upward as lowering $T$. At around $T=T_c/2$, $\alpha\simeq \alpha_{\rm uni}$. This is because the order parameter saturates $S_y\simeq 1$ already at $T=T_c/2$ as shown in Fig.~\ref{fig:Oyz}(b).

\begin{figure*}[t!]
\centering
\includegraphics[width=\textwidth]{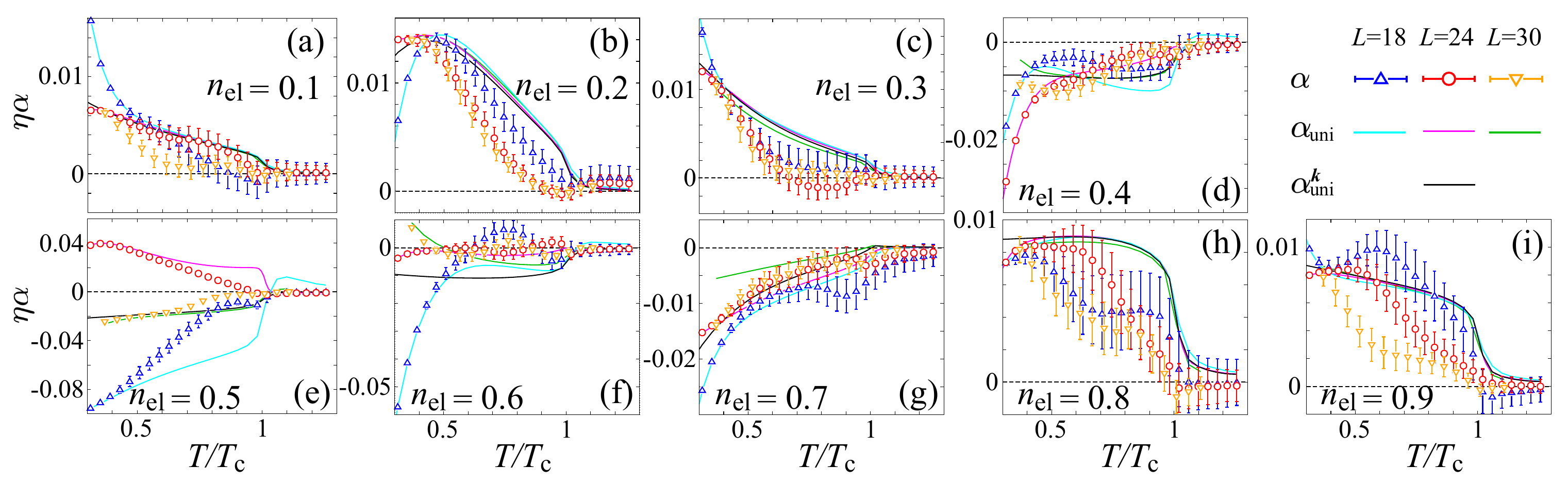}
\caption{$T$ dependence of $\eta\alpha(T,L)$ for $L=18$ (blue triangle), 24 (red circle), and 30 (yellow triangle). (a)--(i) Data for $n_{\rm el}=0.1$--0.9. Lines represent $\alpha_{\rm uni}(T,L)$ for $L=18$ (skyblue), $24$ (magenta), and 30 (green). Black line is $\alpha_{\rm uni}^{\bm{k}}(T,L=24)$, which corresponds to $L\to \infty$ limit of $\alpha_{\rm uni}$ at low $T$ regime. For (d), (e), and (f), there are large finite-size effects both in $\alpha$ and $\alpha_{\rm uni}$.}
\label{fig:alpha_nel}
\end{figure*}

We now discuss the filling $n_{\rm el}$ and the system size dependence of the CIM. Figure \ref{fig:alpha_nel} shows $\eta\alpha(T,L)$, $\eta\alpha_{\rm uni}(T,L)$, and $\eta\alpha_{\rm uni}^{\bm{k}}(T,L)$ as a function of $T$ with several values of $n_{\rm el}$'s. Due to the singularities and/or fine-structure in the DOS as shown in Fig.~\ref{fig:FermiSurface}(a), there are large finite-size effects even in $\chi_{\rm uni}$ for some electron fillings, {\it e.g.,} in Figs.~\ref{fig:alpha_nel}(d) $n_{\rm el}=0.4$, \ref{fig:alpha_nel}(e) $n_{\rm el}=0.5$, and \ref{fig:alpha_nel}(f) $n_{\rm el}=0.6$. Therefore, we cannot carry out any meaningful analysis for these fillings and we concentrate on the data for other values of $n_{\rm el}$ in the following. Firstly, we can observe that there is concrete tendency of suppression of the CIM due to the order parameter fluctuations near $T=T_c$. Apart from the smallest system size data ($L=18$), one finds $|\eta\alpha(T,L)|<|\eta\alpha_{\rm uni}(T,L)|$. Thus, the results discussed in Fig.~\ref{fig:alpha_1} for $n_{\rm el}=0.2$ can be applicable to rather general situations. However, we must still be careful about the finite size effects. For $n_{\rm el}=0.8$ and 0.9, the $\alpha(T,L)$ data for $L=24$ and $L=30$ still noticeably different. For other fillings $n_{\rm el}=0.1$--0.4, and $0.7$, $|\eta\alpha(T,L=24,30)|$'s show concave upward $T-T_c$ dependence near $T_c$ and merge to $\alpha_{\rm uni}^{\bm{k}}$ at low temperature. These are in stark contrast with the convex upward $T$ dependence for $|\eta\alpha_{\rm uni}|$ near $T_c$. See also the $\alpha_{\rm uni}^{(1/8)}$ data Fig.~\ref{fig:alpha_1}. Thus, one can conclude that the order parameter fluctuations suppress the CIM and lead to qualitatively different temperature dependence from one for the order parameter. Namely, the concave upward $T$ dependence can appear near $T_c$.


\section{Discussion}\label{sec:dis}
So far we have shown the numerical results of CIM in chiral quadrupole phase in the distorted kagom\'e structure. We have demonstrated that the order parameter fluctuations strongly suppress the CIM compared with those calculated in the mean-field treatment $\alpha_{\rm uni}$'s. In this section, we will discuss the mechanisms of the suppression in Sec.~\ref{sec:disA}. In Sec. \ref{sec:disB}, we will also point out that the electronic characteristics can lead to concave upward $T$ dependence of the CIM even for the mean-field analysis shown in Fig.~\ref{fig:alpha_nel}. In the last part of this section (Sec.~\ref{sec:disC}), we will comment about future perspective including the extension of the present analysis.

\subsection{Strong suppression of CIM near $T_c$}\label{sec:disA}
\begin{figure}[t!]
\centering
\includegraphics[width=0.48\textwidth]{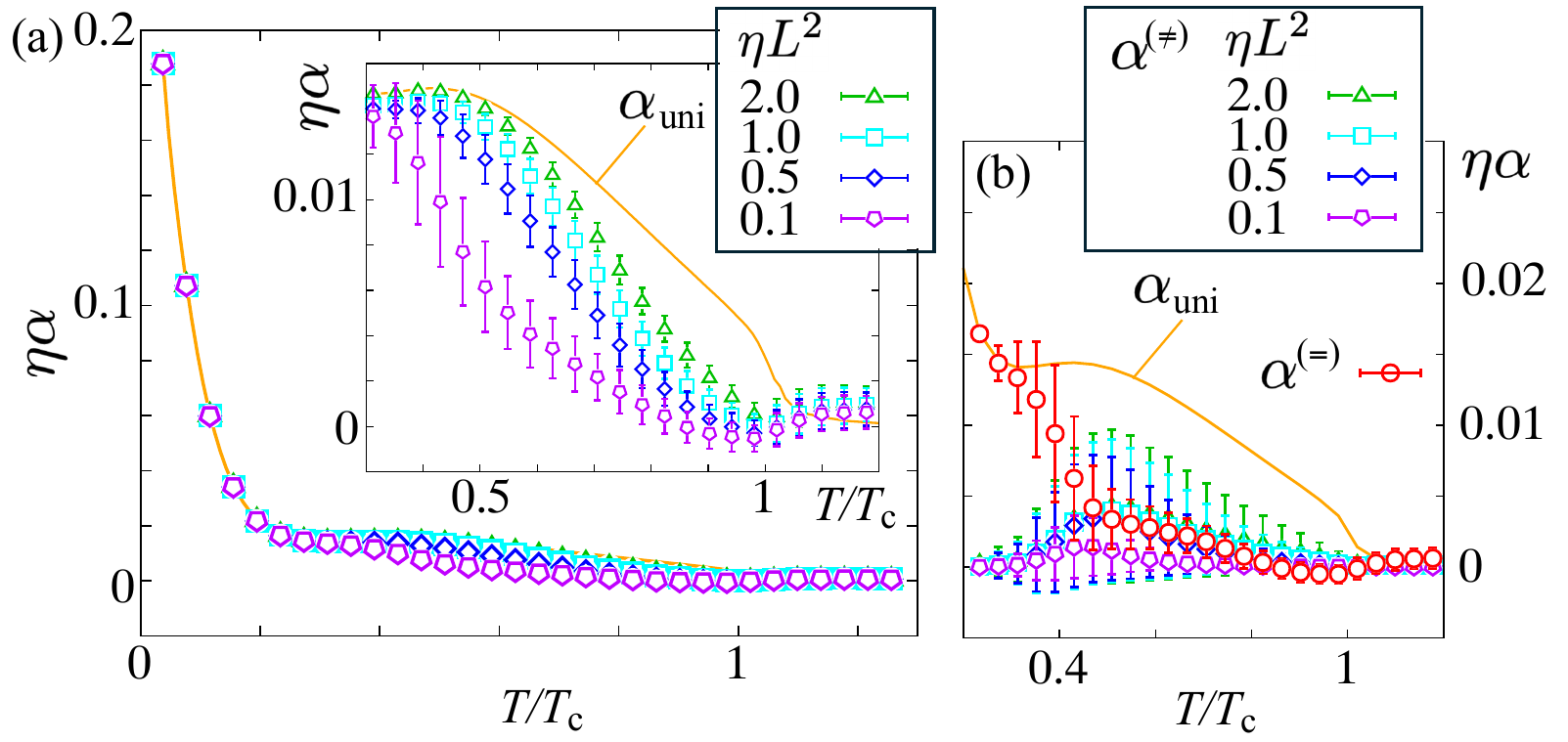}
\caption{(a) $T$ dependence of $\eta \alpha(T,L=24)$ for $\eta L^2=2.0,1.0,0.5$ and $0.1$, and $\eta\alpha_{\rm uni}(T,24)$ (no $\eta$ dependence). Inset: zoom up for $0.3\le T/T_c \le 1.2$. (b)  $T$ dependence of $\eta \alpha^{(=),(\ne)}(T,L=24)$ for $0.2\le T/T_c \le 1.2$. }
\label{fig:alpha_2}
\end{figure}

To unveil the mechanism of the suppression of $\alpha$ near $T=T_c$ due to the order parameter fluctuations, we examine the effects of $\eta$ on the $T$ dependence of CIM. Figure \ref{fig:alpha_2}(a) shows $\eta\alpha(T,L=24)$ for several values of $\eta L^2=0.1,0.5,1.0$, and 2.0 with $n_{\rm el}=0.2$. As shown in the inset, there are noticeable dependence on $\eta$. As $\eta$ becomes large, $\alpha$ approaches $\alpha_{\rm uni}$. This is understood by the microscopic expression (\ref{eq:alpha_Av}). Equation (\ref{eq:alpha_Av}) includes the Lorentzian factor $\mathcal{L}(\epsilon_a-\epsilon_{a'})=\eta/[(\epsilon_a-\epsilon_{a'})^2+\eta^2]$. Thus, for smaller $\eta$, the summation over $a$ and $a'$ is restricted only to the term for $a=a'$ as schematically illustrated in Fig.~\ref{fig:alpha_schematic}(a). In usual situations, it is natural to expect that the matrix element $\sigma_{aa}^i(\bm{r},n) v_{aa}^{j}(\bm{r}'n')$ is smaller in its magnitude than that for the translational symmetric case, where the only diagonal  matrix elements $a=a'\to \bm{k}$ remain finite. In the presence of fluctuations of the order parameters, a part of the weight  originally at the diagonal element spreads to offdiagonal elements with $a\ne a'$, leading to the reduction of the weight at the diagonal element with $a=a'$. Thus, $|\eta \alpha|$ is suppressed for smaller $\eta$. When $\eta$ is large, the width of the Lorentzian function is larger than that for the split energy due to the fluctuations. This effectively leads to recovery of the translational symmetry, since the summation over $a$ and $a'$ contains sufficient numbers of offdiagonal terms, which stems from the diagonal element when the fluctuations are absent. Indeed, the contribution of the diagonal term $\alpha^{(=)}$ for $a=a'$ and that of the offdiagonal term $\alpha^{(\ne)}$  for $a\ne a'$ in Eq.~(\ref{eq:alpha_Av})  show such behavior as a function of $\eta$ in Fig.~\ref{fig:alpha_2}(b). Note that $\eta \alpha^{(=)}$ has no $\eta$ dependence and thus only a set of data is shown (red circle). These results demonstrate that the degeneracy lifting due to the order parameter fluctuations leads to the suppression in the CIM. 


\subsection{Single particle mechanism for enhanced CIM} \label{sec:disB}
\begin{figure}[t!]
\centering
\includegraphics[width=0.48\textwidth]{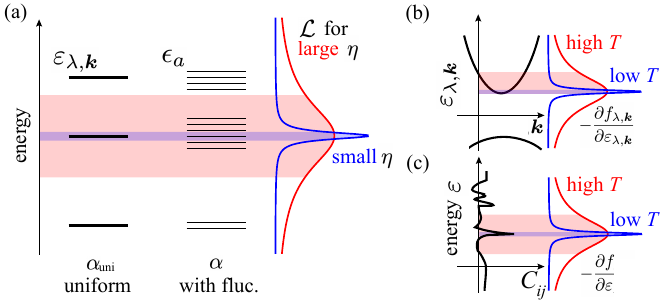}
\caption{Schematic pictures of the single-particle energy spectra and form factors appearing in the expression of CIM [Eqs.~(\ref{eq:alpha_Av}) and (\ref{eq:response})]. Colored (red or blue) area corresponds to the effective width of the integral appearing in Eq.~(\ref{eq:alpha_e}). (a) Comparison between $\alpha_{\rm uni}$ without fluctuations and $\alpha$ with fluctuations. For the energy spectra with the order parameter fluctuations, the degeneracy present in the mean-field treatment (indicated as``uniform'') is lifted. The energy dependence of the Lorentzian form factor ${\mathcal L}$ is also shown for large (red line) and small (blue line) $\eta$. (b) Schematic picture for the case that the band bottom is close to the Fermi energy. Form factor $\partial f/\partial \epsilon$ has a sharp peak at the Fermi energy at low $T$. (c) Schematic picture of $C_{ij}$ with a sharp peak at the Fermi energy and the form factor $\partial f/\partial \epsilon$ at high and low $T$.}
\label{fig:alpha_schematic}
\end{figure}

Here, we discuss mechanism of increasing CIM as lowering $T$ due to particular single-particle properties without order parameter fluctuations discussed so far. Without the order parameter fluctuations, the CIM is given by Eq.~(\ref{eq:response}). This is rewritten in the integral of the single particle energy $\epsilon_{\lambda}$ as 
\begin{align}
\alpha_{ij}(T) &= \frac{1}{\eta} \sum_{\lambda}\int d\epsilon_\lambda C(\epsilon_\lambda)
\left(-\frac{\partial f(\epsilon_{\lambda})}{\partial \epsilon_{\lambda}}\right), \label{eq:alpha_e}
\end{align}
where ``chirality density'' $C(\epsilon_\lambda)$ is given by 
\begin{align}
C_{ij}(\epsilon_\lambda)
&=\frac{1}{N}\sum_{\bm{k}}\delta(\epsilon_\lambda - \epsilon_{\lambda, \bm{k}})\sigma_{i\lambda,{\bk}}v_{j\lambda,{\bm k}}.\label{eq:def_C}
\end{align}
$C_{ij}(\epsilon)$ is the quantity representing the product of the spin and the velocity averaged over the energy shell at $\epsilon$, and thus implicitly includes the DOS at $\epsilon$.  
When a band edge is close to the Fermi energy $(\mu$ at $T=0$) as schematically shown in Fig.~\ref{fig:alpha_schematic}(b), the effective integration region $\propto T$ determined by the factor $-\partial f(\epsilon_\lambda) /\partial \epsilon_{\lambda}$ includes the energy window where $C_{ij}(\epsilon)=0$ since there are no states below the bottom of the band. As lowering $T$, the energy window gradually increases and this leads to increase in $\alpha$. This effect is a part of the reasons for the increase of CIM for $n_{\rm el}=0.2$ in Fig.~\ref{fig:alpha_nel}(b). See also the $T$ dependence of $\mu$ in Fig.~\ref{fig:FermiSurface}(a), where $\mu$ is close to the bottom of the conduction band around the $\Gamma$ point. 

Similar situations occur when the DOS or more directly $C_{ij}(\epsilon)$ shows a sharp peak structure near the Fermi energy as shown in Fig.~\ref{fig:alpha_schematic}(c). For example, let us assume the Lorentzian form of $C_{ij}(\epsilon)$ with the width $\Delta>0$. Then, we can estimate $\eta\alpha_{ij}$ as 
\begin{align}
	\eta\alpha_{ij}(T)&\sim \frac{1}{T} \int d\epsilon \frac{\Delta}{\epsilon^2+\Delta^2} \frac{e^{\epsilon/T}}{(e^{\epsilon/T}+1)^2}\nonumber\\
	&\sim \begin{cases}
		1/T\ \ {\rm for}\ \Delta \ll T\\
		1/\Delta\ \ {\rm for}\ \Delta \gg T.
	\end{cases}
\end{align}
The steep increase in the CIM for $n_{\rm el}=0.7$ in Fig.~\ref{fig:alpha_nel}(g) is due to the sharp peak in $C_{ii}$ as already shown in Fig.~\ref{fig:FermiSurface}(a), where we have denoted $C_{xx}$ as $v_x\sigma_x$. We also note that the signs of $\alpha_{xx}$ in Fig.~\ref{fig:alpha_nel}  correspond to those of $C_{xx}$'s shown in Fig.~\ref{fig:FermiSurface}(a). These are basically unchanged even when the fluctuations are taken into account as shown in Fig.~\ref{fig:alpha_nel}. In the analysis based on the dynamical mean-field theory \cite{Peters2018-ad}, the CIM is enhanced near the coherence temperature below which the Fermi liquid description is valid. This qualitatively agrees with the above mechanism in a sense that there appears a sharp resonance peak near the Fermi level in the DOS in the Kondo system.

\subsection{Future perspectives}\label{sec:disC}
In this paper, we have investigated how the chiral quadrupole order parameter fluctuations affect the CIM and discussed mainly its $T$ dependence below $T_c$. We have taken into account the fluctuations by using snapshot averages of orbital moments, and calculated the CIM by using the Kubo formula [Eqs.~(\ref{eq:response}) and (\ref{eq:alpha_Av})] with phenomenological electron damping factor $\eta$. 
This, however, can be treated in the Monte Carlo method by directly introducing impurity or vacancy in our classical orbital model (\ref{eq:Hclass}). Such analyses need more computational costs but, in principle, tractable. We leave them as a straight forward extension of the present work in future. As for the enhancement of the CIM, electron correlations play an important role as pointed out in 
the previous studies \cite{Fujimoto2007-lb,Peters2018-ad}. When these effects are combined with the order parameter fluctuations discussed in this paper, the temperature dependence of the CIM can have nontrivial behavior and it must be challenging theoretical analysis in future.

As noted in Sec.~\ref{sec:Intro}, the quadrupole order discussed in this paper is expected to be realized in URhSn \cite{Shimizu2020,Ishitobi2025-uq,Harima2023-gd}.  In this material, however, the experimental measurement of CIM has not been carried out so far. We hope the experiments will be done in near future and bring further information about the CIM. 

Since the order parameter considered in this study is nonmagnetic one, it is also interesting to analyze magnetic order parameters which break the inversion symmetry and lead to the magnetoelectric effects, e.g., in magnetic toroidal orders \cite{Hayami2014} and zigzag antiferromagnetic ones \cite{Yanase2014-yl}. For such analysis, the expression of the magnetoelectric response differs from Eqs.~(\ref{eq:response}) and (\ref{eq:alpha_Av}). Thus, the present results do not hold in general. Nevertheless, the method used in this paper is applicable to the system with magnetic degrees of freedom. Comparing the fluctuation effects owing to different kinds of order parameters is important to understand the cross correlated responses in strongly correlated systems.

\section{Summary}\label{sec:sum}
In summary, we have investigated how fluctuations of the chiral quadrupole order parameter influence the current-induced magnetization in a distorted kagom\'e  system. We have elucidated the mechanism behind the suppression of current-induced magnetization and its unusual temperature dependence, which deviates from that of the order parameter. Additionally, we have highlighted the role of certain single-particle electronic properties in affecting the current-induced magnetization and have successfully demonstrated these effects through numerical calculations. We hope that our findings will inspire both theoretical and experimental researchers to further explore cross-correlated phenomena in strongly correlated systems.

\section*{Acknowledgment}
 The authors thank T. Ishitobi, K. Izawa, and H. Kusunose for fruitful discussions.  This work was supported by JSPS KAKENHI (Grant Nos. JP23H04869 and JP23K20824) from the Japan Society for the Promotion of Science.
\vspace{1cm}

\appendix

\section{Hopping Hamiltonian}\label{app:hop}
We summarize the nearest-neighbor hopping Hamiltonian used in the maintext [the first term in Eq.(\ref{eq:conduction})]. 
Consider the nearest-neighbor $(pp\sigma)$ hopping $t_{pp\sigma}>0$ between the site 1 and 2. 
Using the angle $\theta$ representing the distortion, the spin-independent electron hopping from the unit cell $\bm{r}$ and the sublattice $n=2$ to the $\bm{r}$ and $n=1$  in Fig.~\ref{fig:struc}(a) reads as 
\begin{align}
	H^{(1)}_{12}&=t_{pp\sigma}(\cos\theta c_{1x}^\dag+\sin\theta c_{1y}^\dag)
	(\sin\varphi c_{2x}-\cos\varphi c_{2y}),
\end{align}
where $\varphi=\theta-\tfrac{\pi}{6}$. 
Here, we have omitted the unit cell index and the spin $\sigma$. 
Expressing it in the matrix form, this leads to 

\begin{align}
\!\!\!\!H_{12}^{(1)}\!&=\!{\bf c}_1^\dag \hat{\mathcal{H}}_{12}^{(1)}  
{\bf c}_2
\!=\!(c_{1x}^\dag,c_{1y}^\dag) 
\hat{\mathcal{H}}_{12}^{(1)} 
\begin{pmatrix}
c_{2x}\\
c_{2y}
\end{pmatrix}, 
\hat{\mathcal{H}}_{12}^{(1)} \!=\!\hat{\xi}_+^-,\nonumber\\
\!\!\!\!\hat{\xi}_{\rho}^\lambda\!&=\!-\tfrac{t_{pp\sigma}}{2}\Big(\tfrac{1}{2}\hat{1} +
	i\rho \tfrac{\sqrt{3}}{2}\hat{\sigma}^y\!-\!\hat{\sigma}^z\Big)\!-\!
	 t_{pp\sigma} s_\delta\Big(s_\delta\hat{\sigma}^z\!+\!\lambda c_\delta\hat{\sigma}^x\Big),
\end{align}
where $\rho,\lambda=\pm$ and $c_\delta=\cos(\theta-\pi/3)$ and $s_\delta=\sin(\theta-\pi/3)$. 
Note that $\delta\equiv \theta-\pi/3$ is the distortion angle and $\delta=0$ for the regular kagom\'e structure. 
Similarly, the hopping from $\bm{r}-\bm{a}$ and $n=2$ to $\bm{r}$ and $n=1$ in Fig.~\ref{fig:struc}(a) is given by $\hat{\mathcal{H}}_{12}^{(2)}=\hat{\xi}_{+}^+$.
Similar expressions can be derived for the other bonds. 

To summarize, by using the Fourier transform
\begin{align}
	{\bf c}_{\bk n}=
	\begin{pmatrix}
		c_{\bm{k}nx}\\
		c_{\bm{k}ny}
	\end{pmatrix}=
	\frac{1}{\sqrt{N}}\sum_{\bm{r}}e^{i\bk\cdot \bm{r}} \begin{pmatrix}
		c_{\bm{r}nx}\\
		c_{\bm{r}ny}
	\end{pmatrix},
\end{align}
the hopping Hamiltonian  is given as 
\begin{align}
H_{\rm hop}&=\sum_{\bk}({\bf c}_{\bk 1}^\dag,{\bf c}_{\bk 2}^\dag,{\bf c}_{\bk 3}^\dag) 
\mathbf{H}_{1\bk} 
\begin{pmatrix}
{\bf c}_{\bk 1}\\
{\bf c}_{\bk 2}\\
{\bf c}_{\bk 3}
\end{pmatrix},
\end{align}
where the matrix for the nearest-neighbor (NN) hopping $\mathbf{H}_{1\bk}$ is
\begin{align}
\mathbf{H}_{1\bk}=
			\begin{bmatrix}
			0 & \hat{\xi}^-_++\hat{\xi}^+_+e^{ik(-\frac{2\pi}{3})}& 
			\hat{\xi}^-_-+\hat{\xi}^+_-e^{ik(-\frac{\pi}{3})}
			 \\[1mm]
			  & 0 & \hat{\xi}^-_++\hat{\xi}^+_+e^{ik(0)} \\[1mm]
			  &   & 0
		\end{bmatrix}.\label{eq:B:hopMat1}
\end{align}
Here, we have used simplified notation of the wavevector $k(\phi)=\bk\cdot (\cos\phi,\sin\phi)$ and omitted the left bottom part of the matrix. We note that the spin in Eq.~(\ref{eq:B:hopMat1}) is measured in the global coordinates. 

\section{Velocity (current) operators}\label{app:current}
We here list the velocity and equivalently current operators for the conduction electrons. We set the electron charge to $-1$. The local current operator $\bm{j}({\bm{r}},n)$ at ${\bm{r}}_n$, the unit cell position $\bm{r}$ and the sublattice $n$, is defined as 
\begin{align}
	\bm{j}({\bm{r}},n) \equiv \frac{1}{2} \sum_{\sigma=\uparrow,\downarrow}\sum_{\bm{\delta}\in {\rm NN \ bonds}} \!\!\!\!\bm{e}_{\bm{\delta}} \ j_{\sigma,\bm{r}_{n} \to \bm{r}_{n}+\bm{\delta}} ,\label{eq:def_j}
\end{align}
where $\bm{e}_{\bm{\delta}}=\bm{\delta}/|\bm{\delta}|$ with $\bm{\delta}$ being the NN bond vector and $j_{\bm{r}_{n} \to \bm{r}_{n}+\bm{\delta}}$ is the current operator defined on the bond from $\bm{r}_{n}$ to $\bm{r}_{n}+\bm{\delta}$ as 
\begin{align}
	 j_{\sigma,\bm{r}_{n} \to \bm{r}_{n}+\bm{\delta}}
	 =
	 it_{{\bm{r}_{n} \to \bm{r}_{n}+\bm{\delta}}}^{m\gamma,n\zeta}c_{m\gamma\sigma}^\dag(\bm{r}_{n}+\bm{\delta}) c_{n\zeta\sigma}(\bm{r}_{n}) +\ {\rm h.c.}
\end{align}
Here, $c_{n\gamma\sigma}^\dag(\bm{r}_{n}+\bm{\delta})$ is the creation operator at the unit cell $\bm{r}$, the sublattice $n$, and with the orbital $\gamma$ and the spin $\sigma$. The factor $1/2$ in Eq.~(\ref{eq:def_j}) is needed for  preventing the double counting, since any bonds are shared by two sites. The matrix element $t_{{\bm{r}_{n} \to \bm{r}_{n}+\bm{\delta}}}^{m\gamma,n\zeta}$ can be read from Eq.~(\ref{eq:B:hopMat1}). For example, $t_{{\bm{r}_{2} \to \bm{r}_{2}+a_0(\cos(\pi-\theta),\sin(\pi-\theta))}}^{1\gamma,2\zeta}$ with $a_0$ being the distance between the NN sites, is $(\hat{\xi}_+^-)_{\gamma\zeta}$ extracted from the $(1,2)$ component. In the maintext, we have denoted this current operator as $\bm{j}({\bm{r}},n) =(\hat{v}_x(0,\bm{r},n),\hat{v}_y(0,\bm{r},n),\hat{v}_z(0,\bm{r},n))$ in the global coordinates.

%

\end{document}